\documentclass[10pt,twocolumn]{article}
\usepackage{times}
\usepackage[lined,boxed,ruled]{algorithm2e}
\usepackage[dvips]{graphicx}
\usepackage[usenames,dvipsnames]{color}
\usepackage{amssymb}
\usepackage{amsmath}
\usepackage[all,knot]{xy}
\usepackage{apacite}
\begin{document}
\date{}
\title{\textbf{Mapping the Bid Behavior of Conference Referees}}
\author{\textbf{Marko A. Rodriguez$^\star$, Johan Bollen$^\dagger$, Herbert Van de Sompel$^\dagger$}\\
{\small {\it $^\star$ CCS-3: Knowledge \& Information Systems Science Team, Los Alamos National Laboratory}}\\
{\small {\it $^\dagger$ Digital Library Research \& Prototyping Team, Los Alamos National Laboratory}}\\
{\small \{marko, jbollen, herbertv\}@lanl.gov}}
\maketitle{}

\begin{bf}
The peer-review process, in its present form, has been repeatedly criticized. Of the many critiques ranging from publication delays to referee bias, this paper will focus specifically on the issue of how submitted manuscripts are distributed to qualified referees. Unqualified referees, without the proper knowledge of a manuscript's domain, may reject a perfectly valid study or potentially more damaging, unknowingly accept a faulty or fraudulent result. In this paper, referee competence is analyzed with respect to referee bid data collected from the 2005 Joint Conference on Digital Libraries (JCDL). The analysis of the referee bid behavior provides a validation of the intuition that referees are bidding on conference submissions with regards to the subject domain of the submission. Unfortunately, this relationship is not strong and therefore suggests that there exists other factors beyond subject domain that may be influencing referees to bid for particular submissions.\\
\end{bf}

\section{Introduction}

The peer-review process is the most widely accepted method for validating research results within the scientific community. However, its credibility as a valid certification mechanism has come under scrutiny. There exists a rich body of literature that points to many of the inadequacies of the current system \cite{peer:evans1995,peer:munshid2001,peer:bence2004}, but of particular interest to this paper is the issue concerned with ensuring that referees are in fact reviewing manuscripts within their domain of expertise \cite{peer:kassirer1994,peer:eisenhart2002}. There exists a series of stages within the peer-review process that ultimately lead up to a referee review. One of the first and potentially most important stage is the one that attempts to distributed submitted manuscripts to competent referees. Unfortunately, it is difficult to study many of the stages of the peer-review process due to its confidential nature. Therefore, much of the peer-review process, including referee assignment, remains sheltered from the rigors of the scientific method. Fortunately, the program chairs and steering committee of the 2005 Joint Conference on Digital Libraries\footnote{JCDL 2005 is located at: http://www.jcdl2005.org/} (JCDL) has provided the Los Alamos National Laboratory (LANL) Digital Library Research and Prototyping team the referee bid data used for their 2005 conference peer-review process so that referee assignment could be analyzed for this study.\\

\begin{figure}[h!]
			\begin{center}
				\scalebox{1.0}[1.0]{
				\includegraphics[width=0.45\textwidth]{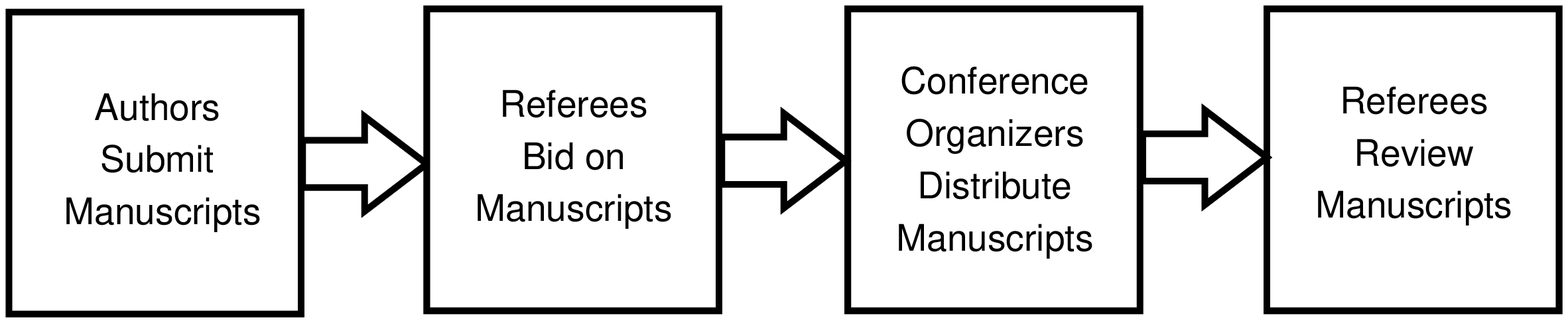}}
			\caption{\label{fig:pipeline} Typical conference review stages}
			\end{center}
		\end{figure}

In conference situations, where there exist a large number of submissions at one particular point in time (near the submission deadline date), conference organizers tend to rely on a pool of pre-selected referees to review the submission archive. The conference organizers require each referee to briefly look over each submission (e.g.~read each submission abstract or ACM classification codes) and place submission bids. A referee bid states the referee's subjective opinion of their level of expertise with regards to a submission. Furthermore, conflict of interest situations are usually identified at this point. Once all the referee bids have been collected, the conference organizers can use any number of the many documented manuscript-to-referee matching algorithms to distribute each submission to a set of competent referees \cite{cheyung:peer1999}. These stages are represented in Figure \ref{fig:pipeline}. The data set provided by the 2005 JCDL program chair does not state which referees reviewed which submission, only the subjective opinion of the referee's level of expertise with respect to each submission.\\

Since conference organizers ask their referees to bid on submissions with regard to their domain of expertise, it is hypothesized that referee bidding is based on two factors: 1) the subject domain of the submission and 2) the expertise of the referee. The validity of this hypothesis is investigated using various statistical techniques that rely on a keyword analysis of submission abstracts and the location of each referee within the greater scientific community's co-authorship network. In short, the analysis demonstrates that the referees of the 2005 JCDL program committee are, in fact, bidding for submissions with respect to the subject domain of the submissions. Unfortunately, the strength of this relationship is not strong enough to conclude that submission subject domain is the only, or even the most significant, factor influencing referee bidding behavior.\\

\section{The 2005 JCDL Bid Data Set}

The JCDL is an international forum that focuses on the technical, practical, and social issues concerning digital libraries. Each year the JCDL hosts a conference to present technical papers, posters, demonstrations, tutorials, etc. that present recent developments in the digital library community.  From June 7$^\mathrm{th}$ to June 10$^\mathrm{th}$ of 2005, the JCDL was held in Denver, Colorado in the United States \cite{sumner:jcdl2005}. The bid data provided by the 2005 JCDL program chair is considered extremely sensitive, therefore careful handling and analysis of this data was the first priority of this research endeavor. All information that is not publicly available from the JCDL website is, to the best of our knowledge, indeterminable from the presented results. Information such as which submissions were rejected is not provided. The names of the referees have been anonymized by assigning each referee a unique random identifier. This section will discuss the bid data provided by the 2005 JCDL program chair as well as the various manipulations necessary to appropriately represent this information for analysis.\\

There were $264$ submissions to the 2005 JCDL. Of those $264$ submissions, $105$ were full technical articles, $77$ were short technical articles, $40$ were posters, $17$ were demonstrations, $4$ were panel talks, $7$ were tutorials, $7$ were workshop talks, and $7$ were doctoral presentations. The JCDL program committee provided the authors a table containing each submission's unique identification number, title, authors, type, and acceptance/rejection status. An example subset of this data is provided in Table \ref{tab:submissioninfo}. The submission titles and authors of those submission that were rejected by the committee have been replaced with the \#\#\# notation in order to protect the privacy of the submitters. Since accepted submissions are freely accessible, information pertaining to accepted publications is provided\footnote{JCDL 2005 proceedings located at: http://www.informatik.uni-trier.de/~ley/db/conf/jcdl/jcdl2005.html}. Furthermore, note that the title and authors have been truncated to ensure that the table fits within the margins of this paper.\\
	
	\begin{table*}[ht!]
		\begin{scriptsize}
		\begin{center}
			\begin{tabular}{|c||c|c|c|c|}
				\hline
				\textbf{sub id} & \textbf{submission title} & \textbf{submission authors} & \textbf{submission type} & \textbf{submission status} \\\hline\hline
				13 & \#\#\# & \#\#\# & Full Technical Article & REJECTED \\\hline
				14 & \#\#\# & \#\#\# & Full Technical Article & REJECTED \\\hline
				15 & Creating an Infrastructure for Collaboration... & R. David Lankes, ... & Short Technical Article & ACCEPTED \\\hline
				16 & Graph-based Text Representation Model... & Hidekazu Nakawatase, ... & Full Techinical Article & ACCEPTED \\\hline
				17 & An Evaluation of Automatic Ontologies... & Aaron Krowne, ... & Full Techinical Article & ACCEPTED \\\hline
			\end{tabular}
		\caption{\label{tab:submissioninfo} Sample of the 2005 JCDL submission data}
		\end{center}
		\end{scriptsize}
	\end{table*}
		
Each referee on the 2005 JCDL program committee was asked to bid on which submissions they wished to review in terms of their expertise in the subject domain of the submission. Therefore, accompanying the submission data table there also exists an associated bid matrix, $\mathbf{B} \in \mathbb{B}^{|S|\times|R|}$, where $S$ is the set of submissions, $R$ is the set of referees, and $\mathbb{B} = \{0,1,2,3,4\}$. It is important to note that $|S| >> |R|$. The rows of the bid matrix refer to the unique id of each of the submissions.  The columns of the bid matrix refer to the referees of the program committee. The matrix entries are the bid values provided by each referee for each submission.  Therefore, $b_{i,j}$ refers to referee $j$'s bid for submission $i$, where $0 \leq b_{i,j} \leq 4$. Table \ref{tab:bidmatrix} is an artificial example of the supplied bid information. Note that the bid values for Table \ref{tab:bidmatrix} were randomly generated and the referee names are not provided. The actual program committee for the JCDL is public information\footnote{JCDL 2005 program committee available at: http://www.jcdl2005.org/progcomm.html}, but their respective bid vectors are not.\\

	\begin{table}[h!]
		\begin{scriptsize}
		\begin{center}
			\begin{tabular}{|c||c|c|c|c|c|}
				\hline
				\textbf{sub/ref} & \textbf{1} & \textbf{2} & \textbf{3} & \textbf{4} & \textbf{5} \\\hline\hline
				\textbf{13} & 1 & 2 & 2 & 3 & 3 \\\hline
				\textbf{14} & 2 & 3 & 2 & 3 & 3 \\\hline
				\textbf{15} & 4 & 2 & 3 & 1 & 1 \\\hline
				\textbf{16} & 3 & 3 & 1 & 2 & 0 \\\hline
				\textbf{17} & 1 & 3 & 2 & 3 & 3 \\\hline
			\end{tabular}
		\caption{\label{tab:bidmatrix} Example bid matrix for each submission for each program committee referee, $\mathbf{B}$}
		\end{center}
		\end{scriptsize}
	\end{table}

The values of the bid matrix, $\mathbf{B}$, are not on an interval scale, but instead are nominal (i.e.~each value symbolizes a particular bid type). Table \ref{tab:bidmeaning} provides the meaning for each of the bid values.\\

	\begin{table}[h!]
		\begin{scriptsize}
		\begin{center}
			\begin{tabular}{|c|l|}
				\hline
				\textbf{bid} & \textbf{meaning of the bid value} \\\hline\hline
				0 & did not provide a bid \\\hline
				1 & expert in the domain of the submission and wants to review \\\hline
				2 & expert in the domain of the submission \\\hline
				3 & not an expert in the domain of the submission \\\hline
				4 & conflict of interest between referee and submission \\\hline
			\end{tabular}
		\caption{\label{tab:bidmeaning} The meaning of the bid values within the bid matrix, $\mathbf{B}'$}
		\end{center}
		\end{scriptsize}
	\end{table}
	
The bid matrix provided by the 2005 JCDL, $\mathbf{B}'$, contains extraneous information such as 'wants to review' ($b=1$) and 'conflict of interest' ($b=4$). Since this study focuses specifically on referee expertise, this information will be discarded.  Therefore, bid categories $1$ and $2$ will be considered the same and bid categories $0$ and $4$ will be considered wildcards. The modified bid matrix used throughout the remainder of this study has the properties of $\mathbf{B}' \in \mathbb{B}'^{|S|\times|R|}$ where $\mathbb{B'} = \{0,1,2\}$. Table \ref{tab:bidmeaning2} has the bid meanings of the modified bid matrix.\\

	\begin{table}[h!]
		\begin{scriptsize}
		\begin{center}
			\begin{tabular}{|c|l|}
				\hline
				\textbf{bid} & \textbf{meaning of the bid value} \\\hline\hline
				0 & unknown expertise (wildcard) \\\hline
				1 & expert in the domain of the submission \\\hline
				2 & not an expert in the domain of the submission \\\hline
			\end{tabular}
		\caption{\label{tab:bidmeaning2} The meaning of the bid values within the modified bid matrix, $\mathbf{B}'$}
		\end{center}
		\end{scriptsize}
	\end{table}

The original artificial bid matrix provided in Table \ref{tab:bidmatrix} is thus transformed into the one shown in Table \ref{tab:bidmatrix2}.\\

	\begin{table}[h!]
		\begin{scriptsize}
		\begin{center}
			\begin{tabular}{|c||c|c|c|c|c|}
				\hline
				\textbf{sub/ref} & \textbf{1} & \textbf{2} & \textbf{3} & \textbf{4} & \textbf{5} \\\hline\hline
				\textbf{13} & 1 & 1 & 1 & 2 & 2 \\\hline
				\textbf{14} & 1 & 2 & 1 & 2 & 2 \\\hline
				\textbf{15} & 0 & 1 & 2 & 1 & 1 \\\hline
				\textbf{16} & 2 & 2 & 1 & 1 & 0 \\\hline
				\textbf{17} & 1 & 2 & 1 & 2 & 2 \\\hline
			\end{tabular}
		\caption{\label{tab:bidmatrix2} Example modified bid matrix for each submission for each program committee referee, $\mathbf{B}'$}
		\end{center}
		\end{scriptsize}
	\end{table}

Of the $264$ submissions, only $118$ of the submissions have actual bid data. This means that $146$ submissions had bids of all $0$. Therefore, only those submissions with a complete set of bid data will be analyzed for the remainder of this study. In addition, of the $76$ program committee members of the JCDL, $11$ members gave no bid information.  No bid information is defined as an individual whose bid vector is all $0$'s. These referees were removed from the analysis.  Finally, since a portion of this analysis is based on co-authorship behavior, those referee committee members not located within the DBLP\footnote{Digital Bibliography and Library Project available at: http://www.informatik.uni-trier.de/~ley/db/} were not included in this study. Of the remaining $65$ referees, $5$ were not in the DBLP. Therefore, the bid matrix as defined for the remainder of this study has $118$ rows (submissions), and $60$ columns (referees), $\mathbf{B}' \in \mathbb{B}'^{118 \times 60}$.\\
	
\section{The Methodology}

\begin{figure}[h!]
			\begin{center}
				\scalebox{1.0}[1.0]{
				\includegraphics[width=0.45\textwidth]{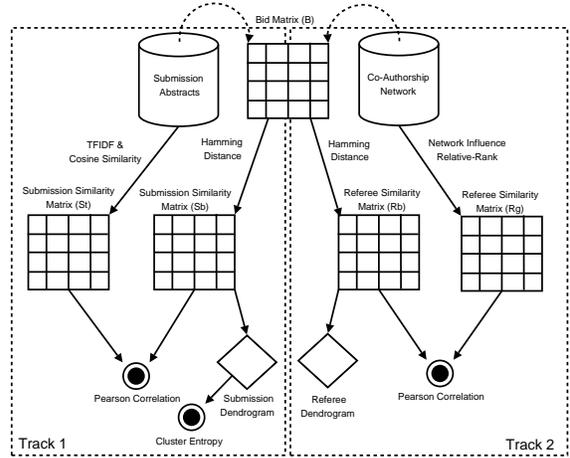}}
			\caption{\label{fig:outline} Experiment outline}
			\end{center}
		\end{figure}

Intuitively, when ignoring conflict of interest situations, referee bidding should be based on two factors: 1) the domain of the submission and 2) the domain of expertise of the referee. Therefore, the referee bid matrix should be the result of each referees analysis of the submission abstracts and the referee's area of expertise (their location in the scientific community's co-authorship network). This idea, which is the hypothesis of this study, is represented by the arced dotted lines at the top of Figure \ref{fig:outline}. To verify or falsify this hypothesis, a collection of statistical techniques are used to determine the relationship between referee bidding and submission subject domain. The two factors of the hypothesis are explored according to Track $1$ and Track $2$ of Figure \ref{fig:outline}.\\  

Track $1$ provides a correlation between two submission similarity matrices.  The first similarity matrix is constructed using referee bid data, $\mathbf{S_b}$, and the second is constructed according to an submission abstract term analysis, $\mathbf{S_t}$ (Section 4). If referees are in fact bidding according to the subject domain of the submissions, then the correlation between $\mathbf{S_b}$ and $\mathbf{S_t}$ should be high. If the correlation is negative, or extremely low, then other factors that may not include submission subject domain are influencing referee bidding. Furthermore, it is possible to cluster submissions according to referee bid behavior. A intra-term analysis of these clusters provide an entropy value for each of the clusters. If the clusters created by referee bidding maintain a low entropy for their highest weighted terms and a low correlation between their term vectors, then it can be argued that referee bidding is driven by submission subject domain.\\

Track $2$ provides the correlation between a referee similarity matrix created according to referee bidding behavior, $\mathbf{R_b}$, and a referee similarity matrix created using a relative rank algorithm within a co-authorship network, $\mathbf{R_g}$ (Section 5). A high correlation means that referees who are similar in expertise, as determined by their place in the co-authorship network, are also bidding similarly. A high correlation would be expected if referee bidding is based solely on submission subject domain. If this correlation is low, then other factors besides submission subject domain are influencing referee bidding. This paper will first explore Track $1$ and then Track $2$.\\

\section{The Bid Matrix and Submission Similarity}

This section will present the Track $1$ analysis represented in Figure \ref{fig:outline}. In order to determine the relationship between referee bidding and submission subject domain, the submissions are related according to the bid behavior of the program committee referees, $\mathbf{S_b}$, and are related according to their abstract term-frequency inverse document-frequency (TFIDF) term weight distributions, $\mathbf{S_t}$ \cite{tfidf:salton1998}. In short, a TFIDF calculation determines the most descriptive words within a document (or document cluster) with respect to the entire document corpus. This section will first discuss the construction of $\mathbf{S_b}$ and then $\mathbf{S_t}$.\\ 

Since the values of the bid matrix, $\mathbf{B}'$, refer to semantic categories and not a gradient scale, a Hamming distance function is used to determine the similarity of any two submissions \cite{hamming:hamming1950}. Hamming distance is defined as the amount of characters that differ between two strings of equal length.  For example, if there exists the strings "2212" and "1212", the Hamming distance is $1$ since only their first characters differ. Given the Hamming distance between two bid vectors, $h(\vec{{b'}_i}, \vec{{b'}_j})$, and the length of a vector, $l=|\vec{b'}_l|$, the similarity between any two submissions is calculated according to Eq. \ref{eq:similarity}. To account for wildcard bids (${b'}_{i,j} = 0$), if any one of the two bid vectors being compared has an entry that contains a $0$, that particular entry on both vectors is ignored and both their vector lengths, $l$, are reduced by $1$.  For example, when comparing the two bid vectors "0121" and "2120", their length, $l$, is $2$ and their Hamming distance, $h$, is $0$ because both their first and last entries are ignored and their second and third entries are equal. Therefore, their similarity is $1$.\\
	
	\begin{equation}
		\label{eq:similarity}
	 		\mathbf{S_b}_{i,j} = \mathbf{S_b}_{j,i} = 1 - \frac{h(\vec{{b'}_i}, \vec{{b'}_j})}{l}
	 \end{equation}
	 
Eq. \ref{eq:similarity} ensures a symmetrical submission similarity matrix, $\mathbf{S_b} \in \mathbb{R}^{|S|\times|S|}$, whose diagonal values are $1.0$. According to the sample bid matrix presented in Table \ref{tab:bidmatrix2}, the submission similarity matrix shown in Table \ref{tab:submissionsimilarity} is constructed using Eq. \ref{eq:similarity}.\\
	
		\begin{table}[h!]
			\begin{scriptsize}
			\begin{center}
				\begin{tabular}{|c||c|c|c|c|c|}
					\hline
					\textbf{id} & \textbf{13} & \textbf{14} & \textbf{15} & \textbf{16} & \textbf{17}\\\hline\hline
					\textbf{13} & 1.0 & 0.8 & 0.25 & 0.25 & 0.8 \\\hline
					\textbf{14} & 0.8 & 1.0 & 0.0 & 0.5 & 1.0 \\\hline
					\textbf{15} & 0.25 & 0.0 & 1.0 & $0.\bar{66}$ & 0.0 \\\hline
					\textbf{16} & 0.25 & 0.5 & $0.\bar{66}$ & 1.0 & 0.5 \\\hline
					\textbf{17} & 0.8 & 1.0 & 0.0 & 0.5 & 1.0 \\\hline
				\end{tabular}
			\caption{\label{tab:submissionsimilarity} Submission similarity determined according to their Hamming distance}
			\end{center}
			\end{scriptsize}
		\end{table}	

\subsection{Submission Similarity and the Dendrogram}

\begin{figure}[h!]
			\begin{center}
				\scalebox{1.0}[1.0]{
				\includegraphics[angle=180,width=0.475\textwidth]{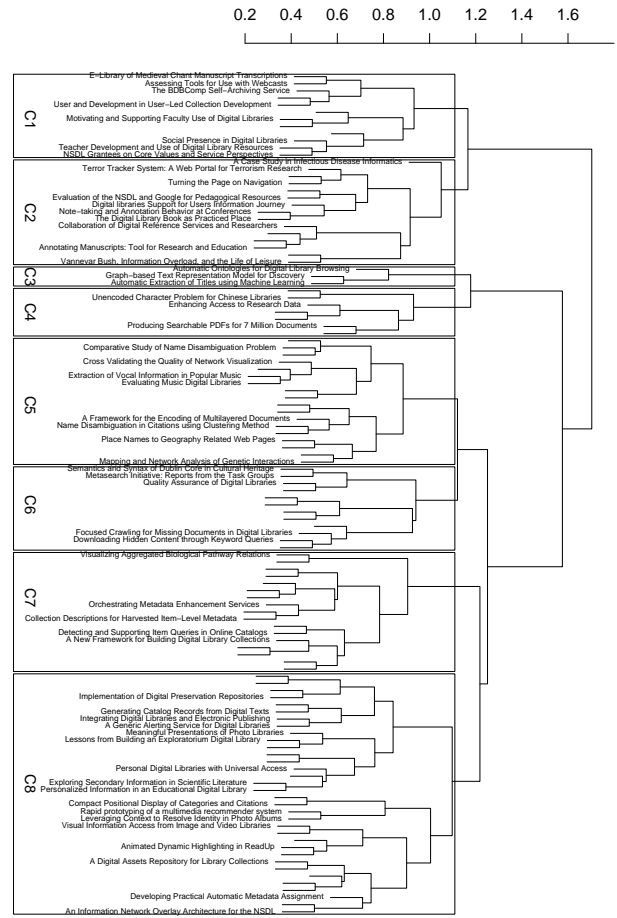}}
			\caption{\label{fig:dendro-submissions} Submission similarity represented according to a hierarchical cluster}
			\end{center}
		\end{figure}

Once a submission similarity matrix, $\mathbf{S_b}$, has been constructed it is possible to hierarchically structure the submissions into a dendrogram in order to visualize the relationship between the various submissions. The submission dendrogram constructed from $\mathbf{S_b}$ is presented in Figure \ref{fig:dendro-submissions}. Note that the titles of the rejected submissions have been left out. Accepted submission titles have been truncated to ensure readability. Furthermore, larger cluster patterns are represented as the $8$ boxed sections and are denoted \textbf{C1} through \textbf{C8}. These clusters were extracted from the dendrogram by setting a threshold on the dendrogram tree height. The threshold, which is $1.1$, was arbitrarily selected to expose enough clusters to make the following analysis interesting.\\

A manual review of the clusters with respect to the submissions they contain demonstrates a congruency between submission topic and referee bidding. To validate this qualitative claim, three statistical techniques are used.  The first involves analyzing the abstracts of the submissions of each of the clusters in order to determine cluster subject domain. The second involves determining the entropy value of each cluster. Clusters that are more strict with respect to a particular subject domain will tend to have a lower entropy. The third technique provides a correlation between a similarity matrix constructed from the cosine similarity of the TFIDF term weight vectors of each submission, $\mathbf{S_t}$, and the matrix constructed from the referee bid behavior, $\mathbf{S_b}$. This final correlation provides a single quantitative value expressing the relationship between submission subject domain and referee bidding.\\  
		
\subsection{Entropy in the Submission Clusters}

The $8$ major clusters of the submission dendrogram derived according to referee bid data can be validated as meaningful categorizations by analyzing the terms of the submission abstracts. This requires that all submission abstracts be parsed to determine the full collection of keywords across all submission abstracts. Each abstract is processed by removing stop words and then applying the Porter stemming algorithm \cite{stemmer:porter2005}. These two processes remove overly frequent words (i.e.~the, and, it) and perform suffix stripping (i.e.~computer and computation stem to comput), respectively. Each time a particular term in the full collection of keywords is used in an abstract of one of the cluster submissions, the term frequency for that term in that cluster is incremented by $1$. An example feature vector is provided in Table \ref{tab:featurevector}. For example, for all the submissions in cluster $3$, the term {\it built} was used $7$ times.\\

		\begin{table}[h!]
			\begin{scriptsize}
			\begin{center}
				\begin{tabular}{|c||c|c|c|c|}
					\hline
					\textbf{cluster/term} & \textbf{browser} & \textbf{built} & \textbf{bureau} & \textbf{bush} \\\hline\hline
					\textbf{3} & 3 & 7 & 3 & 1 \\\hline
					\textbf{4} & 4 & 3 & 2 & 0 \\\hline
					\textbf{5} & 1 & 0 & 1 & 0 \\\hline
				\end{tabular}
			\caption{\label{tab:featurevector} Cluster feature vectors of the keywords in the submission abstracts}
			\end{center}
			\end{scriptsize}
		\end{table}	

For each cluster $i$ it is possible to determine how specific a particular term $j$ is to that cluster according to Eq. \ref{eq:tfidf} where $\mathrm{freq}(i,j)$ is the frequency of term $j$ in cluster $i$, $n(i)$ is the total number of terms in cluster $i$, $N$ is the number of clusters (which is always $8$ for this experiment), and $n_c(j)$ is the number of clusters for which term $j$ appears \cite{tfidf:salton1998}.\\

	\begin{equation}
		\label{eq:tfidf}
	 		\mathrm{tfidf}(i,j) = \frac{\mathrm{freq}(i,j)}{n(i)} \times \mathrm{log}_{10}\left(\frac{N}{n_c(j)}\right)
	\end{equation}

The higher the TFIDF weight for term $j$ in cluster $i$, the more specific term $j$ is to the cluster $i$ and therefore the more suited it is as a description of the cluster's subject domain. The following table presents the TFIDF calculations for the sample feature vector presented in Table \ref{tab:featurevector}.\\	
	
		\begin{table}[h!]
			\begin{scriptsize}
			\begin{center}
				\begin{tabular}{|c||c|c|c|c|}
					\hline
					\textbf{cluster/term} & \textbf{browser} & \textbf{built} & \textbf{bureau} & \textbf{bush} \\\hline\hline
					\textbf{3} & 0.00 & 0.08 & 0.00 & 0.03 \\\hline
					\textbf{4} & 0.00 & 0.05 & 0.00 & 0.00 \\\hline
					\textbf{5} & 0.00 & 0.00 & 0.00 & 0.00 \\\hline
				\end{tabular}
			\caption{\label{tab:weightvector} Cluster TFIDF term weight vectors of the keywords in the submission abstracts}
			\end{center}
			\end{scriptsize}
		\end{table}	
		
In order to determine the subject domain of each of the $8$ clusters, the top 10 TFIDF weighted terms were extracted. Table \ref{tab:clusterterms} provides these terms ordered by their TFIDF weight where term $1$ has a higher weight than term $2$.\\ 
		
		\begin{table*}[ht!]
			\begin{scriptsize}
			\begin{center}
				\begin{tabular}{|c||c|c|c|c|c|c|c|c|}
					\hline
					& \textbf{C1} & \textbf{C2} & \textbf{C3} & \textbf{C4} & \textbf{C5} & \textbf{C6} & \textbf{C7} & \textbf{C8} \\\hline\hline
					\textbf{1} & webcast & behaviour & extract & patent & name & hidden & ecl & photo \\\hline
					\textbf{2} & cyberinfrastructur &	drew & powerpoint & tobacco &	surrog & crawler & morf & preserv \\\hline
					\textbf{3} & ncknow & note & handel & invent & music &	subschema & relev & video \\\hline
					\textbf{4} & interview & overload &	mainli & determin &	disambigu &	flora & item-level & european \\\hline
					\textbf{5} & teacher & engag & train & hidden & segment &	ontos & network & alert \\\hline
					\textbf{6} & descriptor & factor & graph & control & candid & queri & citat & dark \\\hline
					\textbf{7} & faculti &	gather & step & american & network & expans & circleview & region \\\hline
					\textbf{8} & lesson & school & weight & chemic & tempor & homepag & dlii & busi \\\hline
					\textbf{9} & survei & teamsearch & algebra & compani & citat & plant & extract & addit \\\hline
					\textbf{10} & transcript & visualis & basi & searchabl & genet & reusabl & meta-inform & mobil \\\hline
				\end{tabular}
			\caption{\label{tab:clusterterms} Top 10 terms for the 8 clusters defined in Figure \ref{fig:dendro-submissions}}
			\end{center}
			\end{scriptsize}
		\end{table*}

The term weight distributions derived from the TFIDF calculation of the cluster abstracts can now be represented according to their internal cluster information content. Internal cluster information content can be calculated using the standard entropy equation as defined according to its information theoretic sense \cite{entropy:shannon1948}. The lower the entropy, the more specialized, or focused, the cluster. The higher the entropy, the less specialized. Since clusters vary in size, the entropy for a cluster is calculated only for the top $10$ term weights presented in Table {\ref{tab:clusterterms}.  Furthermore, since an entropy calculation is defined for a probability distribution, Eq. \ref{eq:normalize} normalizes the top $10$ term weights.\\

	\begin{equation}
		\label{eq:normalize}
	 		\mathrm{tfidf}'(i,j) = \frac{\mathrm{tfidf}(i,j)}{\sum_{k=0}^{k<10} \mathrm{tfidf}(i,k)}
	\end{equation}

The entropy of a cluster is then calculated over the probability distribution as described by Eq. \ref{eq:entropy}, where $H(i)$ is the entropy for cluster $i$.\\

	\begin{equation}
		\label{eq:entropy}
	 		H(i) = - \sum_{j=0}^{j<10} \mathrm{tfidf}'(i,j) \; \mathrm{log}_2(\mathrm{tfidf}'(i,j))
	\end{equation}
	
The entropy values for the $8$ clusters of the dendrogram presented in Figure \ref{fig:dendro-submissions} are presented in Table \ref{tab:clusterentropy}.\\
	
		\begin{table}[h!]
			\begin{scriptsize}
			\begin{center}
				\begin{tabular}{|c|c|c|}
					\hline
					\textbf{cluster} & \textbf{entropy} \\\hline\hline
					\textbf{1} & 3.2668 \\\hline
					\textbf{2} & 3.2840 \\\hline
					\textbf{3} & 3.2148 \\\hline
					\textbf{4} & 3.2213 \\\hline
					\textbf{5} & 3.2025 \\\hline
					\textbf{6} & 3.2281 \\\hline
					\textbf{7} & 3.2610 \\\hline
					\textbf{8} & 3.2442 \\\hline
				\end{tabular}
			\caption{\label{tab:clusterentropy} Entropy values for the 8 clusters defined in Figure \ref{fig:dendro-submissions}}
			\end{center}
			\end{scriptsize}
		\end{table}

It is interesting to note that \textbf{C5}, the lowest entropy cluster, is composed mainly of submissions associated with name disambiguation and music in digital-library research.  On the other hand, the highest entropy cluster \textbf{C2} has a mix of more unrelated submissions ranging from digital libraries in educational settings to infectious diseases and terrorism. Figure \ref{fig:cluster-distribution1} and \ref{fig:cluster-distribution2} present the distribution of the term weights for the top $10$ terms of the $8$ clusters.  The steeper the distribution tail, the lower the cluster entropy and therefore the more focused the cluster is towards its higher weighted terms. The analysis of the terms for each cluster points to a qualitative relationship between referee bidding and submission subject domain.\\

		\begin{figure}[h!]
			\begin{center}
				\scalebox{1.0}[1.0]{
				\includegraphics[angle=-90,width=0.48\textwidth]{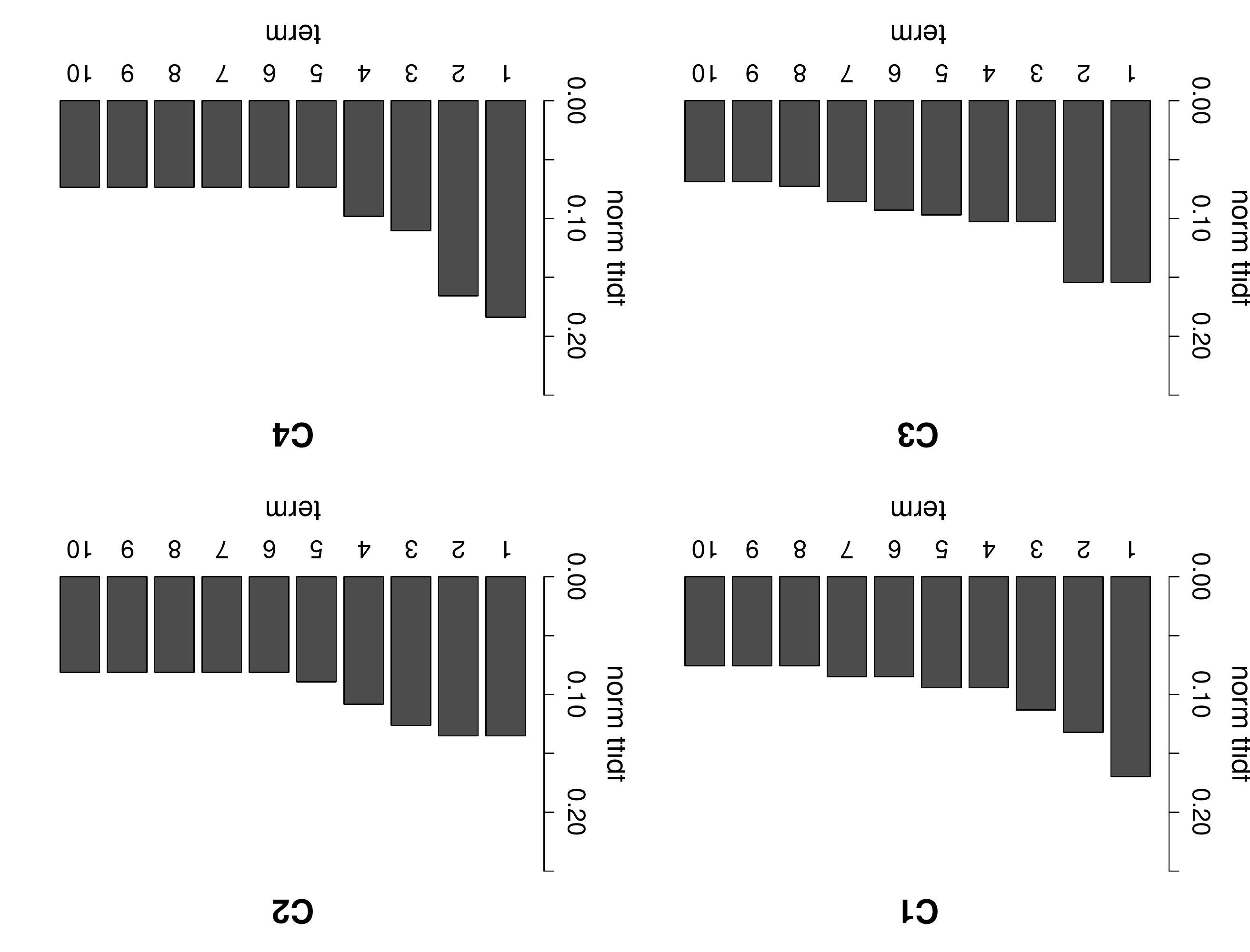}}
			\caption{\label{fig:cluster-distribution1} Normalized TFIDF weights for the 10 terms of Table \ref{tab:clusterterms} for the clusters 1 through 4 defined in Figure \ref{fig:dendro-submissions}}
			\end{center}
		\end{figure}
		
		\begin{figure}[h!]
			\begin{center}
				\scalebox{1.0}[1.0]{
				\includegraphics[angle=-90,width=0.48\textwidth]{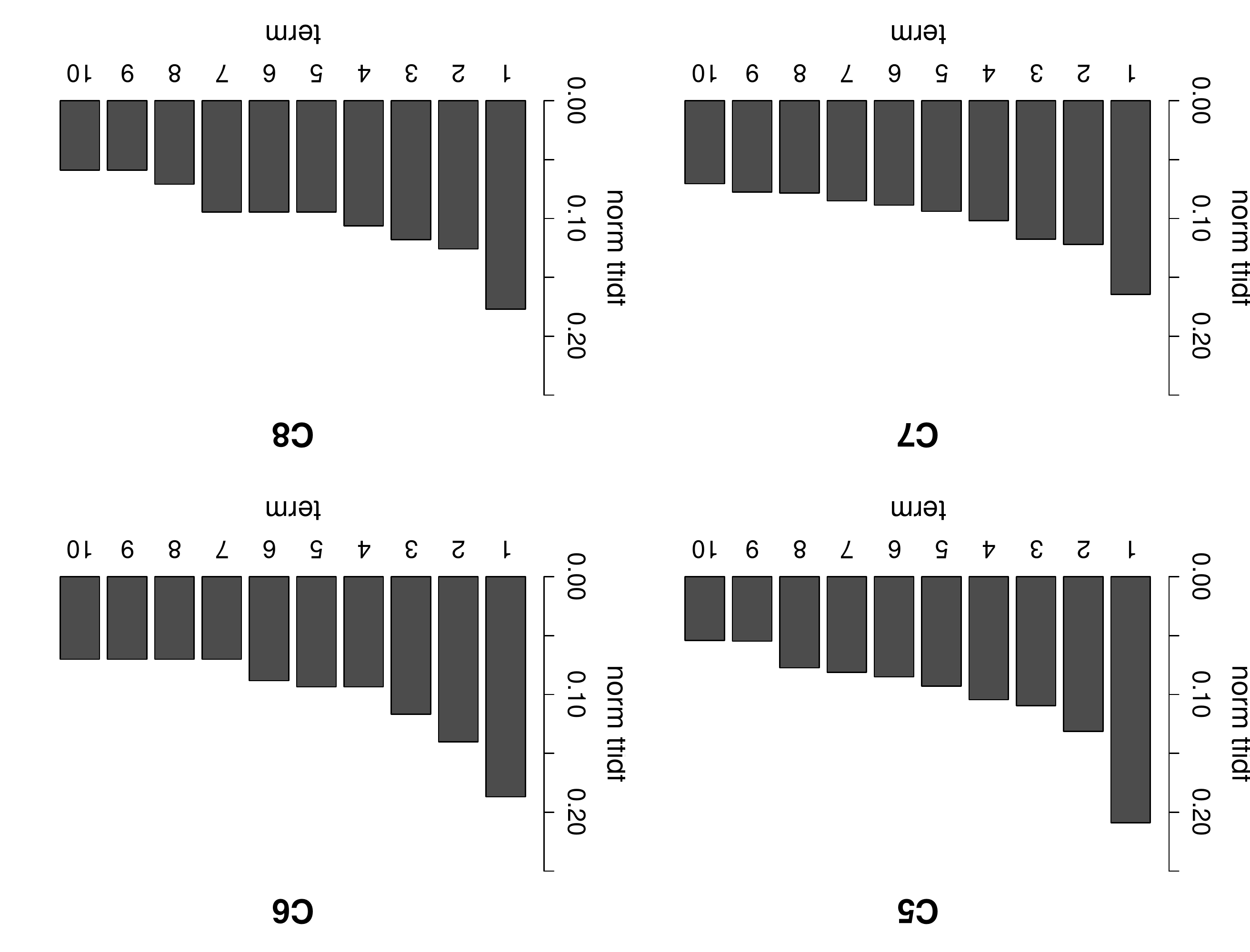}}
			\caption{\label{fig:cluster-distribution2} Normalized TFIDF weights for the 10 terms of Table \ref{tab:clusterterms} for the clusters 5 through 8 defined in Figure \ref{fig:dendro-submissions}}
			\end{center}
		\end{figure}

A more quantitative validation can be determined when the TFIDF term weight vectors of the $8$ clusters are compared using a Pearson correlation.  The correlations are performed on the cluster's TFIDF term weight vectors which contain the entire abstract dictionary, $D$, where $|D| = 2121$. Table \ref{tab:clustercorrelation} provides the Pearson correlations for each cluster comparison. What is noticeable from Table \ref{tab:clustercorrelation} is that all the correlations are less than $\pm0.1$. The fact that the clusters, which are organized by referee bidding, yield very low correlations between their TFIDF term weight vectors means that the clusters are well separated according to their term distributions. If these correlations were high, then it would be difficult to claim that the clusters are organized according to subject domain and thus referee bidding would not be related to submission subject domain. Since the correlations are all less that $\pm0.1$, this confirms the hypothesis that there does exist a relationship between the bidding behavior of the conference referees and the subject domain of the submission abstracts.  The next section will further explore the strength of this relationship.\\
		
		\begin{table*}[ht!]
			\begin{scriptsize}
			\begin{center}
				\begin{tabular}{|c||c|c|c|c|c|c|c|c|}
					\hline
					& \textbf{C1} & \textbf{C2} & \textbf{C3} & \textbf{C4} & \textbf{C5} & \textbf{C6} & \textbf{C7} & \textbf{C8} \\\hline\hline
					\textbf{C1} & 1.0    & -0.0318  & -0.0396 & -0.0631 & -0.0616 & -0.0477 & -0.0720 & -0.0527 \\\hline
					\textbf{C2} & -0.0318 & 1.0     & -0.0214 & -0.0639 & -0.0392 & -0.0502 & -0.0323 & -0.0804 \\\hline
					\textbf{C3} & -0.0396 & -0.0214 & 1.0     & -0.0343 & 0.01064 & 0.02349 & 0.04540 & -0.0321 \\\hline
					\textbf{C4} & -0.0631 & -0.0639 & -0.0343 & 1.0     & -0.0437 & -0.0531 & -0.0540 & -0.0850 \\\hline
					\textbf{C5} & -0.0616 & -0.0392 & 0.01064 & -0.0437 & 1.0     & -0.0410 & 0.03928 & -0.0684 \\\hline
					\textbf{C6} & -0.0477 & -0.0502 & 0.02349 & -0.0531 & -0.0410 & 1.0     & -0.0156 & -0.0780 \\\hline
					\textbf{C7} & -0.0720 & -0.0323 & 0.04540 & -0.0540 & 0.03928 & -0.0156 & 1.0     & -0.0687 \\\hline
					\textbf{C8} & -0.0527 & -0.0804 & -0.0321 & -0.0850 & -0.0684 & -0.0780 & -0.0687 & 1.0    \\\hline
				\end{tabular}
			\caption{\label{tab:clustercorrelation} Pearson correlations for the $2121$ TFIDF term weights of the 8 clusters defined in Figure \ref{fig:dendro-submissions}}
			\end{center}
			\end{scriptsize}
		\end{table*}

\subsection{Cosine Similarity Correlation} 

To further quantify the relationship between referee bidding and a submission's subject domain, it is possible to correlate the relationship between submissions based on referee bidding, on the one hand, and the relationship between submissions based on their TFIDF term weight vectors, on the other. This requires the construction of the similarity matrix $\mathbf{S_t}$, which denotes the cosine similarity between every submission with respect to their complete TFIDF term weight vector. This means that each term in the abstracts of each submission is analyzed according to the TFIDF equation presented in Eq. \ref{eq:tfidf}. This results in a matrix $\mathbf{T} \in \mathbb{R}^{|S|\times|D|}$ where $|S|$ is the size of the submission archive and $|D|$ is the size of the full collection of terms of all abstracts in the submission archive. For this particular experiment $\mathbf{T}$ is therefore defined as $\mathbf{T} \in \mathbb{R}^{118 \times 2121}$. The TFIDF term weight vector of each submission can be compared against every other submission's TFIDF term weight vector using the standard cosine similarity function presented in Eq. \ref{eq:cosinesimilarity}, where $\vec{t}_i$ is the TFIDF term weight vector for submission $i$. This equation guarantees a symmetrical matrix with a diagonal of $1.0$.\\

	\begin{equation}
		\label{eq:cosinesimilarity}
	 		\mathbf{S_t}_{i,j} = \mathbf{S_t}_{j,i} = \frac{\vec{t}_i \cdot \vec{t}_j}{\left\|\vec{t}_i\right\| \cdot \left\|\vec{t}_j\right\|}
	\end{equation}

The correlation between $\mathbf{S_t}$ and $\mathbf{S_b}$ can now be calculated. With 13,922 degrees of freedom and a $p$-value $< 2.2^-16$, the Pearson correlation was determined to be $0.357$. This means that submissions categorized according to a TFIDF analysis of their abstracts and submissions categorized according to the referee bid behavior are in fact positively correlated, though not strongly. Therefore, it can be concluded that there are other factors besides submission subject domain that influence referee bid behavior.\\

\begin{figure}[h!]
	\centerline{
		\xymatrix@R=0.1pt{
			*+[F-]{df = 13922,\;p < 2.2^{-16},\;r = 0.357}\\
		}
 	}
\end{figure}	

\section{The Bid Matrix and Referee Similarity}

This section will overview the experiment as described by Track $2$ of Figure \ref{fig:outline}. If referees are deemed similar in expertise, as determined by their relative location to one another within the scientific community's co-authorship network, then similar referees should be bidding similarly. To test this hypothesis, two referee similarity matrices are created. The first referee similarity matrix, $\mathbf{R_b} \in \mathbb{R}^{|R|\times|R|}$, is constructed from the transpose of the modified bid matrix, ${\mathbf{B}'}^T$.  Each referee is compared to each other referee with respect to their bidding behavior.  Based on the transpose of the artificial data from Table \ref{tab:bidmatrix}, the same similarity equation used to construct the submission similarity matrix, Eq. \ref{eq:similarity}, can be used to construct the referee similarity matrix presented in Table \ref{tab:refereesimilarity}. The next section will present a dendrorgam of $\mathbf{R_b}$ before discussing the second referee similarity matrix, $\mathbf{R_g}$.\\

		\begin{table}[h!]
			\begin{scriptsize}
			\begin{center}
				\begin{tabular}{|c||c|c|c|c|c|}
					\hline
					\textbf{ref} & \textbf{1} & \textbf{2} & \textbf{3} & \textbf{4} & \textbf{5}\\\hline\hline
					\textbf{1} & 1.0 & 0.5 & 0.75 & 0.0 & 0.0 \\\hline
					\textbf{2} & 0.5 & 1.0 & 0.2 & 0.4 & 0.75 \\\hline
					\textbf{3} & 0.75 & 0.2 & 1.0 & 0.2 & 0.0 \\\hline
					\textbf{4} & 0.0 & 0.4 & 0.2 & 1.0 & 1.0 \\\hline
					\textbf{5} & 0.0 & 0.75 & 0.0 & 1.0 & 1.0 \\\hline
				\end{tabular}
			\caption{\label{tab:refereesimilarity} Referee similarity determined according to their Hamming distance}
			\end{center}
			\end{scriptsize}
		\end{table}	

\subsection{Referee Similarity and the Dendrogram}

Given the referee similarity matrix, $\mathbf{R_b}$, the dendrogram in Figure \ref{fig:dendro-referees} can be constructed. Unfortunately, due to privacy issues, the referee names are not provided. What is noticeable from the dendrogram is the collection of nearly identical referees on the upper branch. When reviewing the modified bid matrix, $\mathbf{B}'$, it becomes apparent that $19$ of the referees stated themselves to be expert in the domain of every submission (excluding their wildcard bids).\\

		\begin{figure}[h!]
			\begin{center}
				\scalebox{1.0}[1.0]{
				\includegraphics[angle=180,width=0.45\textwidth]{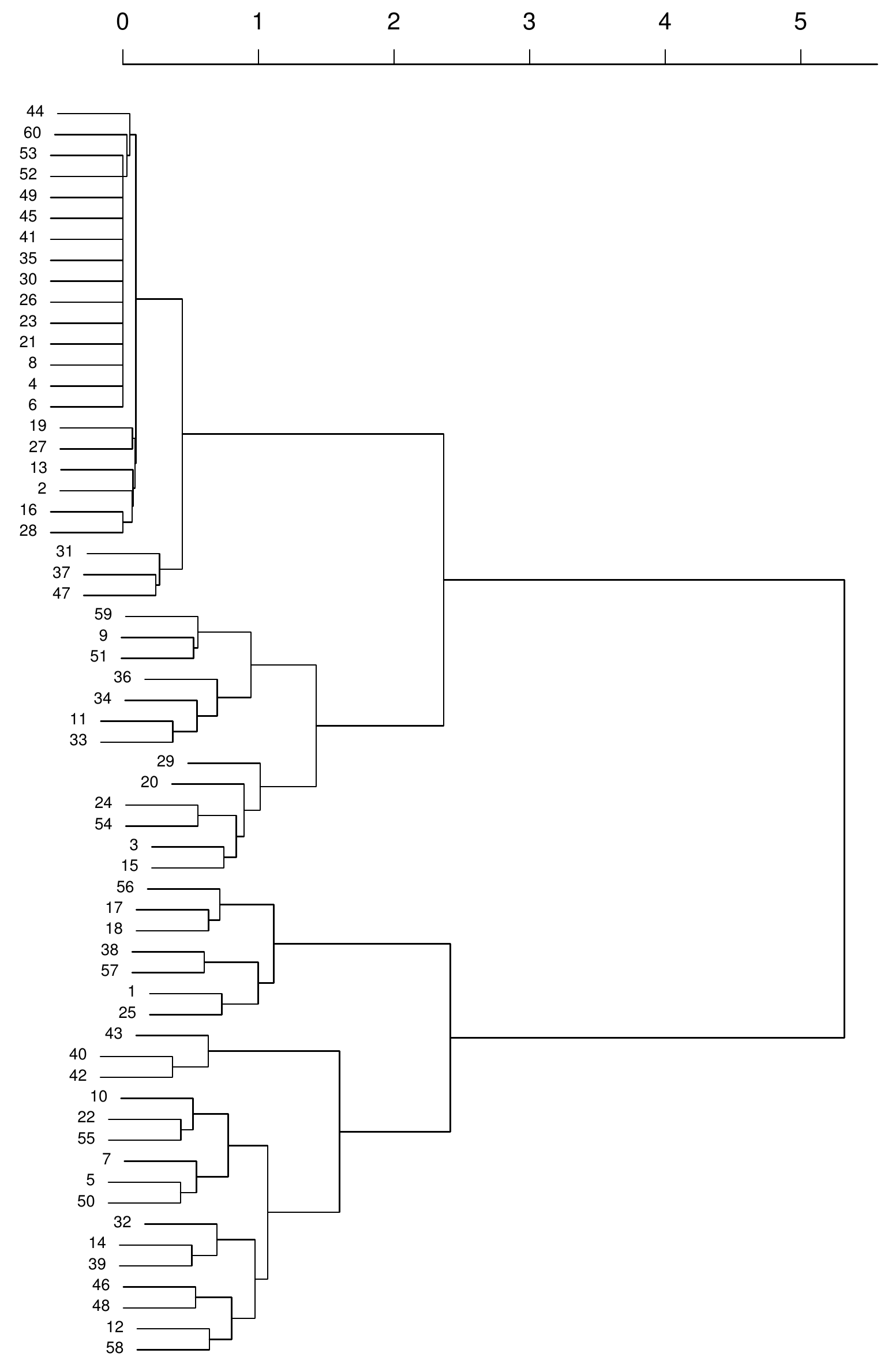}}
			\caption{\label{fig:dendro-referees} Referee similarity represented according to a hierarchical cluster}
			\end{center}
		\end{figure}

\subsection{Relative-Rank Correlation}

In order to provide a quantitative evaluation of the similarity of referees with respect to their bidding behavior and their domain of expertise, a relative-rank algorithm within a co-authorship network is computed to determine referee similarity. It has been widely accepted that co-authorship networks represent the relationship of individuals with respect to their domain of expertise \cite{newman:coauthor2004}. The relative-rank algorithm will determine the similarity of each referee with respect to each other referee as defined by their relative location to one another within the greater scientific community's co-authorship network. The similarity of the referees as determined by their relative-rank, $\mathbf{R_g}$, and their similarity as determined by their bid behavior, $\mathbf{R_b}$, can then be correlated.  A high correlation means that referees of similar expertise are bidding in a similar manner. A low correlation means that referees of similar expertise are not bidding in a similar manner. The co-authorship network, $G$, used for this experiment was constructed from the DBLP database as of October 2005. The DBLP co-authorship network has 284,082 nodes (authors) and 2,167,018 edges (co-authorship relationships). This section will first formalize the co-authorship network data structure and relative-rank algorithm before discussing the results.\\

A co-authorship network is defined by a graph composed of nodes that represent authors and edges that represent a joint publication. Therefore, a co-authorship network is represented by the tuple $G=(N,E,W)$, where $N$ is the set of authors in the network, $E$ is the set of edges relating the various authors, and $W$ is the set of weights associated with the strength of tie between any two collaborating authors. Any edge, $e_{i,j}$, connects two authors, $n_i$ and $n_j$, with a respective weight of $w_{i,j}$. Furthermore, $E \subseteq N \times N$ and $|E| = |W|$.  The edge weight between any two authors is determined by Eq. \ref{eq:edgeweight}, where the summation is over the set of all manuscripts registered with the DBLP, $M$, expressing a collaboration between authors $n_i$ and $n_j$, and the function $A(m)$ returns the total number of authors for manuscript $m$, where $m \in M$ and $w_{i,j} \in \mathbb{R}^+$ \cite{coauth:liu2005,newman:science2001}.\\

	\begin{equation}
		\label{eq:edgeweight}
	 		w_{i,j} = w_{j,i} = \sum_{\forall m \in M \;\mathrm{authored \; by}\; i,j} \frac{1}{A(m)-1}
	 \end{equation}

To provide the reader with an understanding of the relationship between the 2005 JCDL program committee members, a subset of the DBLP co-authorship network which contains the program committee's co-authorship relationships is presented in Figure \ref{fig:refereegraph}. Note that this network was not constructed using referee bid data, but from information that is publicly available through the DBLP database. Furthermore, the co-authorship edge weights have been left out to improve readability.\\ 

		\begin{figure}[ht!]
			\begin{center}
				\scalebox{1.0}[1.0]{
				\includegraphics[angle=90,width=0.45\textwidth,height=0.55\textheight]{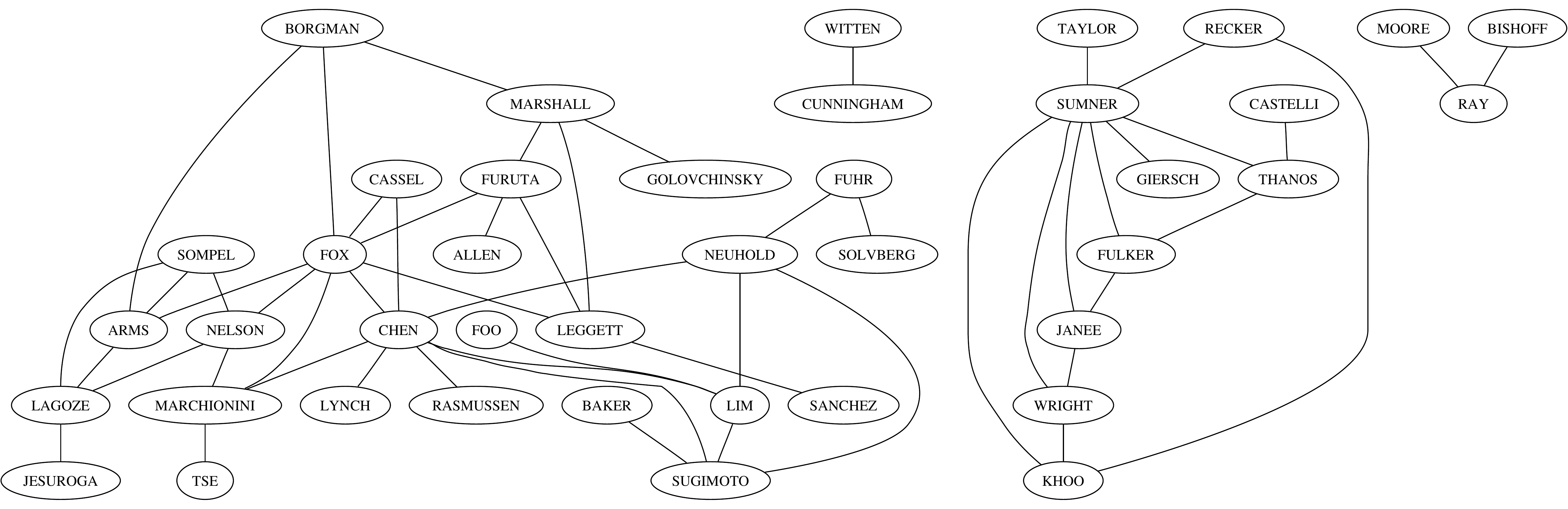}}
			\caption{\label{fig:refereegraph} Subset of the DBLP co-authorship network containing only connected JCDL referees}
			\end{center}
		\end{figure}
	 
The final analysis to be performed is to rank each of the $60$ referees relative to one another so as to construct $\mathbf{R}_g \in \mathbb{R}^{|R| \times |R|}$, Eq. \ref{eq:authorsimilarity}. For each referee in the JCDL program committee that provided valid bid data and is located in the DBLP, a similarity value to every other member in the committee was computed using a relative-rank algorithm (sometimes called a 'personalized' rank) \cite{partpage:rodriguez2005,markov:white2003} within the DBLP co-authorship network. Since $G$ is a weighted graph, the ranking algorithm actually used in this experiment is the weighted relative-rank implementation described in \cite{partpage:rodriguez2005}.\\

	\begin{equation}
		\label{eq:authorsimilarity}
			\mathbf{R_g} = \left(
				\begin{array}[pos]{ccc}
					\mathbf{R_g}_{R_1, R_1} & \cdots & \mathbf{R_g}_{R_1,R_{|R|}} \\
					\vdots & \ddots & \vdots \\
					\mathbf{R_g}_{R_{|R|}, R_1} & \cdots & \mathbf{R_g}_{{R_{|R|},R_{|R|}}} \\
				\end{array}
				\right)
	 \end{equation}

An example of relative-ranking is as follows. Given a network such as the one displayed in Figure \ref{fig:refereegraph}, the relative-rank algorithm would rank {\it FOX} more strongly to {\it NELSON} than to {\it RAY} since there exists a clique relationship between {\it FOX}, {\it NELSON}, and their co-authors. This network structure does not exist between {\it FOX} and {\it RAY}. Since co-authorship networks relate individuals with respect to similar domains of expertise, the conclusion to be drawn is that the stronger ranking of {\it FOX} to {\it NELSON} implies that {\it FOX} is more related by expertise to {\it NELSON} than he is to {\it RAY}. A simplified version of the  pseudo-code for constructing $\mathbf{R_g}$, Eq. \ref{eq:authorsimilarity}, is presented in Algorithm 1. For a more indepth, and formal, review of relative-rank algorithms for network analysis, refer to \cite{partpage:rodriguez2005,markov:white2003}.\\

	\incmargin{1cm}
	\restylealgo{boxed}
	\linesnumbered
	\begin{algorithm}[h!]
	\begin{scriptsize}
	\Indp
			\ForEach{$(n_l \in R)$}{
					\ForEach{$(n_j \in R)$}{
						$\mathbf{R_g}_{{n_l},{n_j}} = \mathrm{rank}(n_l,n_j)$\;
					}
				}
		\label{alg:authorsim}
		\caption{Constructing the referee similarity matrix $\mathbf{R_g}$}
	\end{scriptsize}
	\end{algorithm}
	\decmargin{1em}
	 
Given $\mathbf{R_b}$ and $\mathbf{R_g}$, with 3,598 degrees of freedom and a $p$-value $< 2.2^{-16}$, the Pearson correlation was calculated to be $0.220$. The positive correlation indicates that referees are bidding with respect to their domain of expertise, but the low correlation again hints that there may be other factors contributing to referee bidding.\\

\begin{figure}[h!]
	\centerline{
		\xymatrix@R=0.1pt{
			*+[F-]{df = 3598,\;p < 2.2^{-16},\;r = 0.220}\\
		}
 	}
\end{figure}

\section{Conclusion}

This paper provided an exploration of the bidding behavior of the 2005 JCDL program committee.  The various analysis techniques used demonstrate that the 2005 JCDL program committee did, in fact, bid for conference submissions with respect to the subject domain of the submission. On the other hand, the strength of this relationship is low and therefore demonstrates that other factors may be involved in referee bidding. One such factor seems to be referee fatigue. With $146$ submissions having no bid data and with $19$ referees stating themselves to be an expert in the domain of all submissions, human-driven referee bidding in conference settings may not be the most optimal technique for performing conference peer-review. Since bidding is the preliminary component of the manuscript-to-referee matching algorithm, sloppy bidding can have dramatic effects on which referees actually review which submissions, Figure \ref{fig:pipeline2}. In general, the stages that follow from the inclusion of noisy data in the peer-review chain can severely effect the quality of the peer-review process. It is speculated that referee fatigue not only influences the bidding and manuscript dissemination stages of the review cycle, but potentially more damaging, fatigued referees could be rejecting acceptable manuscripts or accepting fraudulent or faulty manuscripts in the review stage.\\

		\begin{figure}[h!]
			\begin{center}
				\scalebox{1.0}[1.0]{
				\includegraphics[width=0.45\textwidth]{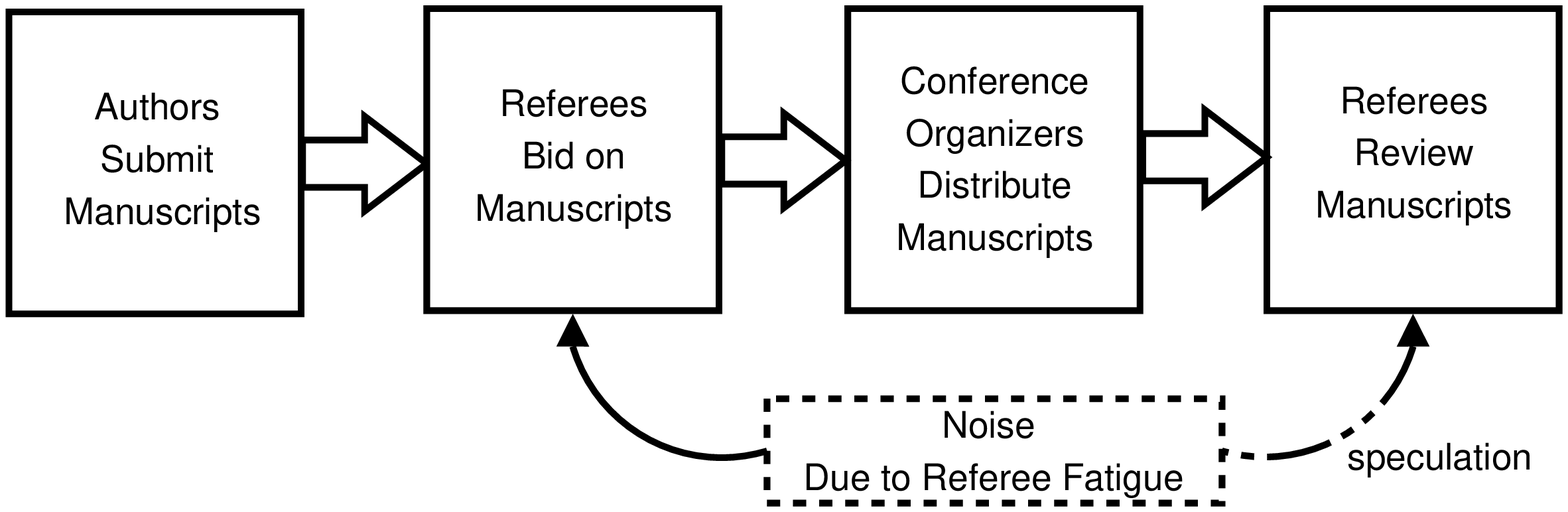}}
			\caption{\label{fig:pipeline2} Human-factor noise in review stages}
			\end{center}
		\end{figure}

When the $19$ referees that provided no variation in their bid vector (referees that stated themselves to be experts in the domain of all submissions) are removed from the analysis, the correlation between $\mathbf{S_b}$ and $\mathbf{S_g}$ was determined to be $0.361$ (originally $0.357$). Likewise, for $\mathbf{R_b}$ and $\mathbf{R_g}$ the correlation was determined to be $0.383$ (originally $0.220$). Though these are both higher correlations, they still are not strong correlations. It can only be concluded that, contrary to the hypothesis, submission subject domain and referee expertise are not the only factors involved in the referee bidding process. Future work in this area will focus on expanding this study's hypothesis in order to develop a mathematical model of the factors influencing referee bidding. Furthermore, an application of this methodology to bid data from other conferences can help to provide a broader perspective (and confirmation) of the factors influencing submission bidding in the peer-review process.\\

\section{Acknowledgments}
This research could only have been conducted with the help of the 2005 JCDL program chairs (Mary Marlino, Tamara Sumner, Frank Shipman) and steering committee (Erich Neuhold). Furthermore, the DBLP has once again provided the authors of this paper a thorough dataset for exploring the structure of science. The authors would like to thank the producers of R Statistics, GraphViz, Dia, and LaTeX for their freely available software. Finally, Xiaoming Liu and Karin Verspoor contributed by reviewing drafts of this manuscript. This research was financially supported by the Los Alamos National Laboratory.\\

\bibliographystyle{apacite}
\bibliography{markobib}

\end{document}